\documentclass[11pt,twoside]{article}
\pdfoutput=1
\usepackage{asp2006}
\usepackage{graphicx}

\begin{document}
\title{Observations of Polarized Dust Emission at Far-infrared through Millimeter Wavelengths}
\author{John E. Vaillancourt}
\affil{Division of Physics, Mathematics, \& Astronomy,\\California Institute of Technology,\\MS 320-47, 1200 E. California Blvd., Pasadena, CA 91125, USA}

\begin{abstract}
  Interstellar polarization at far-infrared through millimeter
  wavelengths ($\lambda \sim 0.1$ -- 1 mm) is primarily due to thermal
  emission from dust grains aligned with magnetic fields.  This
  mechanism has led to studies of magnetic fields in a variety of
  celestial sources, as well as the physical characteristics of the
  dust grains and their interaction with the field.  Observations have
  covered a diverse array of sources, from entire galaxies to
  molecular clouds and proto-stellar disks.  Maps have been generated
  on a wide range of angular scales, from surveys covering large
  fractions of the sky, down to those with arcsecond spatial resolution.
  Additionally, the increasing availability of observations at
  multiple wavelengths in this band allows empirical tests of models of
  grain alignment and cloud structure.
  I review some of the recent work in this field, emphasizing
  comparisons of observations on multiple spatial scales and at
  multiple wavelengths.
\end{abstract}


\section{Introduction}

Dust grains in both diffuse and dense phases of the interstellar
medium (ISM) are preferentially aligned by local magnetic fields
\citep[e.g.,][]{alexreview2}.  The resulting net polarization is
observed across a wide range of wavelengths.  At visible and
near-infrared wavelengths transmitted starlight has a net polarization
due to dichroic extinction from the net alignment of aspherical
grains.  At longer far-infrared (FIR) and submillimeter (SMM)
wavelengths ($\lambda \sim 0.1$ -- 1 mm) the polarized radiation is
dominated by thermal emission from similar grains. These different
polarization mechanisms result in orthogonal polarizations with
respect to the aligning field, perpendicular to the field in the case
of emission but parallel in the case of absorption and extinction.

Polarimetric observations in the FIR--MM regime cover angular scales
ranging from $\sim$$1\arcsec$ to all-sky.  The goal at all scales is
to determine the structure of the magnetic field.  Previously,
observations were limited to $\sim 0.3$ -- a few arcminutes
\citep[e.g.,][]{archivestokes,hertzarchive,scubaarchive}; the limits
being set by diffraction at the small end and the size of the largest
detector-array field of view (FOV) at the high end.  Interferometers
are now able to regularly achieve $\sim 1\arcsec$ resolution while
large beamsizes coupled with efficient scan strategies have made it
easier to maps regions larger than a single focal plane FOV
\citep[e.g.,][]{page2007,bicep2008}.

The increasingly wide range of wavelengths used to observe at each of
these scales now allows studies of the polarization spectrum.  Just as
in the case of total emission, a well sampled spectrum is used to
place constraints on models of molecular cloud structure as well as
the grain alignment mechanism and its efficiency
\citep[e.g.,][]{whittet08,pspec}.  However, this spectrum is only
sparsely sampled and further data are needed to better characterize the
shape of the spectrum and how it changes across different physical
environments.

\section{Magnetic Fields}

\subsection{Galactic-Scale Fields}

The polarization of background starlight at optical wavelengths
towards thousands of stars has done an excellent job of tracing
large-scale magnetic fields in the Galaxy (e.g.,
\citealt{mf1970,heilescat}; \citealt*{berdyugin04}).  These
observations trace the field only in diffuse regions of the
ISM where optical photons can penetrate ($A_V <$ few magnitudes). The
inferred field is in general parallel to the Galactic plane.
However, in denser regions of the ISM as traced by FIR--SMM
polarimetry it is unclear whether there exists such a correlation with
the Galactic plane \citep[e.g.,][]{tenerife,li06,paris}.

Further information on the Galactic magnetic field comes from MM-wave
observations intended to study Galactic emission as a foreground
contaminant to the cosmic microwave background (CMB\@). An all-sky
survey at 94\,GHz ($\lambda = 3.2$\,mm) performed by the WMAP
satellite finds a mean magnetic field parallel to the plane
\citep{kogut2007,page2007,hinshaw2008}, in good agreement with optical
polarimetry. WMAP's spatial resolution for polarization detections at
94\,GHz is poor ($\sim 4\deg$) compared to other existing optical and
FIR--SMM observations.  For this reason, and the fact that its long
wavelength is more sensitive to colder dust, WMAP is likely sampling
much more of the magnetic field in the diffuse ISM than are shorter
wavelength FIR--SMM observations.
That is, it is sampling more of the dust also seen at optical
wavelengths than in the FIR--SMM\@.  
Future CMB polarization experiments will have increased spatial
resolution ($\sim$ several arcminutes) and extend to wavelengths as
short as $850\,\micron$ \citep{bicep2008,tauber04,aumont08}.

\subsection{Intermediate Scales}

At the smallest angular scales gravitational collapse, turbulent
motions, and other effects are likely to distort the magnetic field
direction with respect to any regularity that may be imposed upon it
by the larger ambient Galactic field.  A number of authors have tried
to compare the field direction inferred from starlight polarization
with that from SMM polarimetry.  The success has been limited due to
the very few locations where such a comparison can be undertaken, and
the difficulty of estimating and removing foreground polarization from
the starlight measurements
\citep{w3,dasth,li06,poidevin06,poidevin08}.

The limited sensitivity of the current generation of SMM polarimeters
does not allow magnetic fields to be mapped with high spatial
resolution towards precisely the same areas as existing starlight
polarization data in diffuse areas of the ISM\@.  However, sensitivity
to low-surface brightness clouds can be increased by either binning
maps made at high-spatial resolution, or designing an instrument with
intrinsically large beams.  Such measurements can be compared to
higher resolution observations to investigate how the magnetic field
structure changes across multiple cloud scales.  For example, in the
Galactic center, SMM polarimetry on degree scales ($5\arcmin$
resolution) reveals a toroidal field, parallel to the plane
\citep{novak03}.  This is in contrast to the existence of non-thermal
(synchrotron emitting) filaments with long axes perpendicular to the
Galactic plane (\citealt*{yz84,yz04}; \citealt{nord04}), which has
been taken as evidence of a poloidal field within $20\deg$ of the
Galactic center.  However, on smaller scales (arcminutes) toroidal
fields are observed in the densest clouds but poloidal fields in the
less dense regions \citep{chuss03}.  This apparently discrepant
structure has been interpreted as evidence of a globally poloidal
Galactic center field which has been sheared into a toroidal field due
to rotation of the dense material about the center.

\begin{figure}[!tb]
\centering
\includegraphics[height=0.50\textheight]{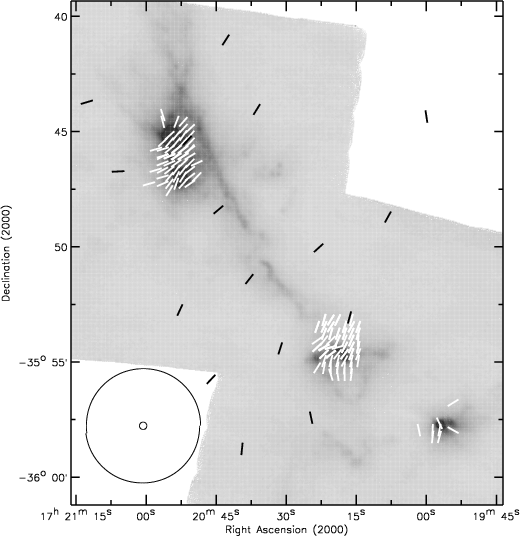}
\caption{The position angle direction of the magnetic field in
  NGC\,6334 is inferred from polarization observations at
  $350\,\micron$ (white vectors; \citealt{hertzarchive}) and
  $450\,\micron$ (black vectors; \citealt{li06}). The circles at
  lower-right show the relative beam sizes of the polarization
  measurements ($350\,\micron = 20\arcsec$; $450\,\micron =
  5\arcmin$). The grayscale intensity was obtained with the
  $350\,\micron$ camera SHARC-II at the Caltech Submillimeter
  Observatory.  The intensity is shown with a logarithmic-stretch and
  has a spatial resolution of $10\arcsec$ (courtesy C. D. Dowell).}
\label{fig-6334}
\end{figure}

A similar comparison can be made in the filamentary molecular cloud
NGC\,6334 (Fig.\ \ref{fig-6334}).  The large-scale magnetic field
($5\arcmin$ resolution) is mostly perpendicular to the long axis of
the filament.  This filament contains 4 dense cores whose polarization
has been mapped at higher spatial resolution ($20\arcsec$).  The field
in the cores exhibits clear similarities to the large-scale field at
the maximum extent of the high-resolution data but also shows clear
deviations from this field as one moves closer to the core centers.

\subsection{Cloud and Star Formation}

In models of magnetically regulated cloud and star-formation,
molecular clouds are divided into two classes
\citep[e.g.,][]{mouschovias76,tomisaka88}.  In the supercritical case,
the cloud mass is large enough that gravitational collapse can proceed
even against the outward force of magnetic pressure. 
In the subcritical case the  magnetic field prevents compression
perpendicular to the field lines, and the cloud can only collapse
parallel to the field. In this case, one might expect clouds to be
flattened parallel to the field.

\begin{figure}[!tb]
\centering
\includegraphics[width=2.0in]{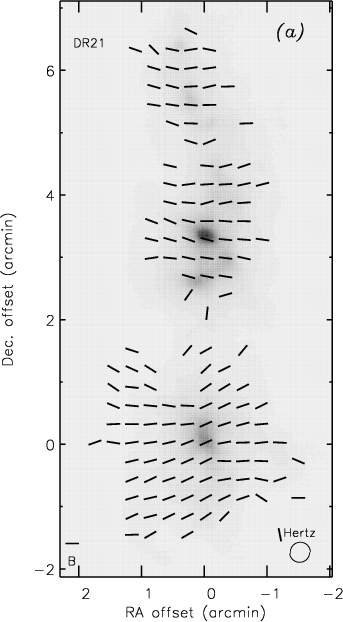}
\includegraphics[width=3.0in]{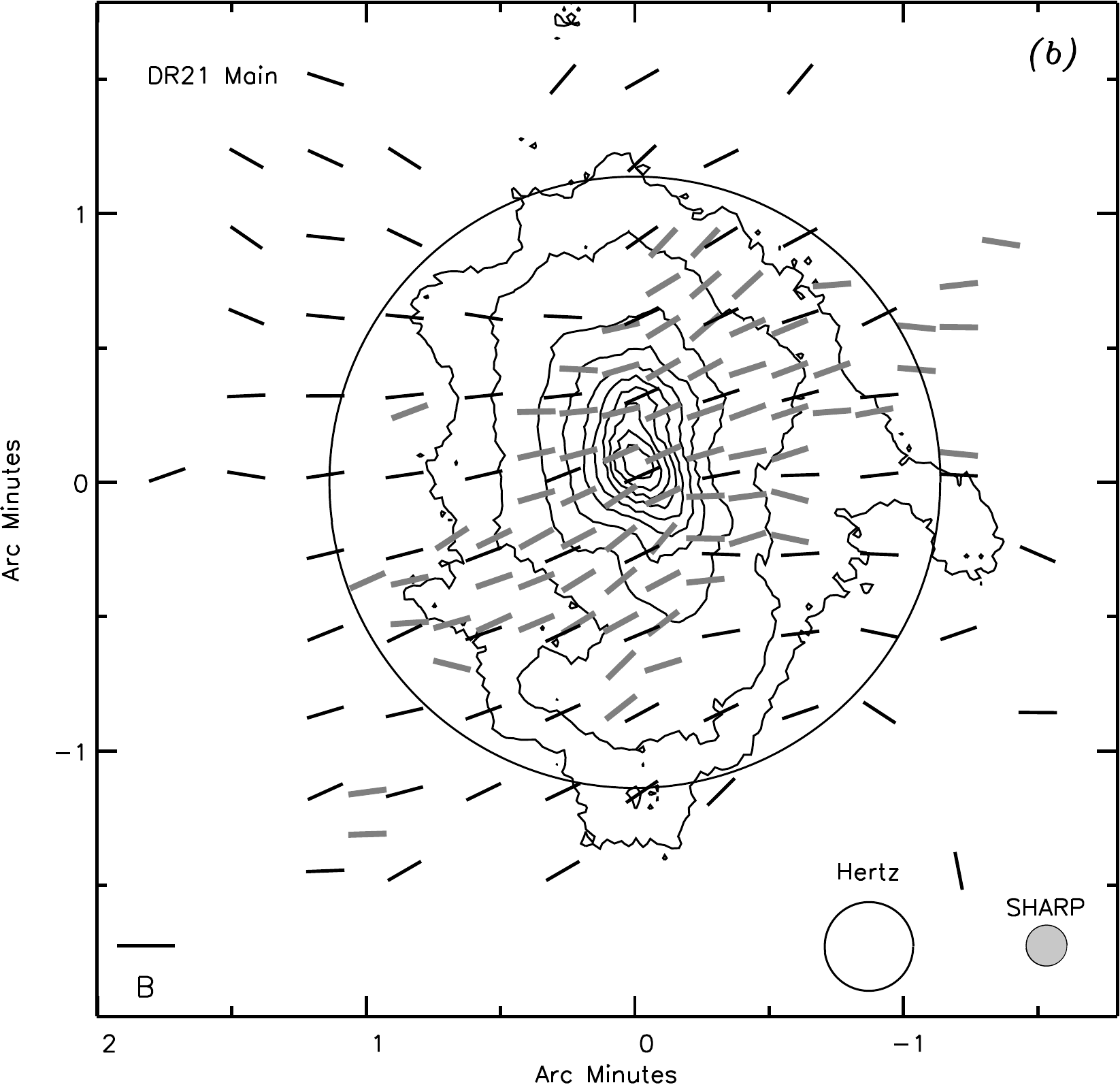}
\caption{Flux and polarization maps of DR\,21 at $350\,\micron$
  \citep{kirby2008}. (\emph{a})~Inferred magnetic field direction in
  the DR\,21 filament at $20\arcsec$ resolution
  \citep{hertzarchive}. North-to-south this map includes the cores
  DR21\,OH\,FIR2, DR21\,OH, and DR21\,Main. (\emph{b}) DR21\,Main
  only, with data from both Hertz (black vectors) and SHARP
  ($10\arcsec$ resolution; gray vectors). The large circle has a
  radius of 1\,pc and indicates the distance from the core beyond
  which the cloud is prevented from undergoing gravitational collapse
  due to magnetic pressure; the cloud should collapse only within the
  circle.  Intensity maps from SHARC-2 (courtesy C.~D. Dowell).}
 \label{fig-dr21}
\end{figure}

Observational examples of this geometry include the clouds DR\,21
(Fig.\ \ref{fig-dr21}\emph{a}; \citealt{kirby2008}), OMC-1
\citep{dasth}, and OMC-3 \citep*{mwf01}.  Tassis et al.\ (in
prep.)\ have studied how the mean projected magnetic field correlates
with cloud elongation and polarization position angle using a large
collection of 100 and 350 $\micron$ polarization data
\citep{archivestokes,hertzarchive}.  A preliminary analysis suggests
that a model in which the field is perpendicular to the long axis of
the clouds is favored over other orientations, even when accounting
for projection effects.  However, other studies suggest that the
elongation directions are random with respect to the field direction
\citep*{glenn99}.  Drawing strong conclusions from the existing work
is difficult, as these data sets are limited to fewer than 20 clouds
and few have large aspect ratios.

The magnetic pressure that maintains subcritical clouds can do so
only by acting on charged particles.  As a result, dense cores may
evolve from a subcritical to supercritical state as neutrals diffuse
through the field (ambipolar diffusion).  The subsequent gravitational
collapse pulls the field along, resulting in the classical
``hour-glass'' field morphology (e.g.\
\citealt*{shu87,vallee2003}). ``Pinched'' fields in OMC-1
\citep{dasth}, DR\,21\,Main (Fig.\ \ref{fig-dr21}\emph{b};
\citealt{kirby2008}), and NGC\,1333 IRAS 4A \citep*{girart06} are
observational examples of this geometry.

The observations of OMC-1 and NGC\,1333 do not cover sufficient
spatial extent such that the magnetic field is observed to merge back
into the larger-scale Galactic field in which the clouds are
embedded. However, in the case of DR21\,Main, \citet{kirby2008} has
argued that the orientation of magnetic field vectors support a model
in which the central portions of the cloud are undergoing
gravitational collapse, but the outer portions of the cloud are
supported by magnetic pressure, retaining the field direction in the
ambient ISM\@.  Figure \ref{fig-dr21}\emph{b} shows $350\,\micron$
intensity and polarization maps of this cloud. By estimating the cloud
mass and magnetic field strength as a function of distance from the
core, \citet{kirby2008} finds that the gravitational and magnetic
energies are balanced at a distance of approximately 1\,pc from the
central core.  As a result the cloud is expected to collapse
gravitational on smaller scales, but remain supported at larger
scales.  This is consistent with the observation that the magnetic
field vectors are more nearly parallel to the ambient field (traced by
the filament in Fig.\ \ref{fig-dr21}\emph{a}) at the larger scales.

To further test theories of magnetically regulated star and cloud
formation polarization maps need to be extended into lower column
density regions of these clouds.  For example, while the observation
of NGC\,1333 IRAS\,4A \citep{girart06} has excellent spatial
resolution ($\sim 1\arcsec$) the measurements do not clearly connect
to the larger-scale ambient medium (the interferometric measurements
are not sensitive to the more extended emission).
Thus these polarization measurements will need to be supplemented by
single dish measurements (e.g., \citealt{sharpao,bastien08}) in order
to connect such small- and intermediate-scale observations.

\section{Polarization Spectra}

\subsection{Empirical Results}

At visible wavelengths, much has been inferred about dust grain physical
properties from spectropolarimetry (e.g.,
\citealt{whittet01,whittet08,whittet04,whittet_here,shape}). For example, we
know that large grains (radii $>0.1\,\micron$) are more efficient
polarizers than small grains (radii $<0.01\,\micron$),
silicate grains are better aligned than graphite grains,
and
the shape of aligned grains is more oblate (disc-like) than prolate
(needle-like).

\begin{figure}[!tb]
\centering
\includegraphics[width=0.52\textwidth]{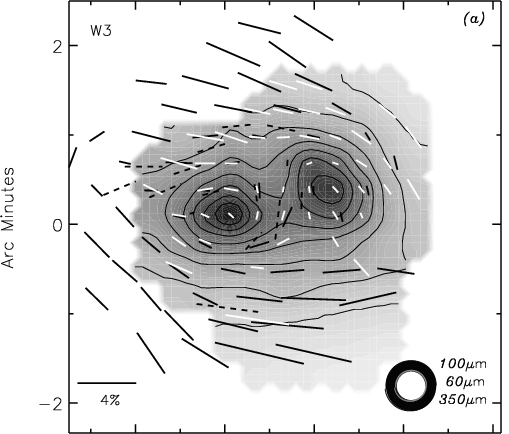}
\includegraphics[width=0.52\textwidth]{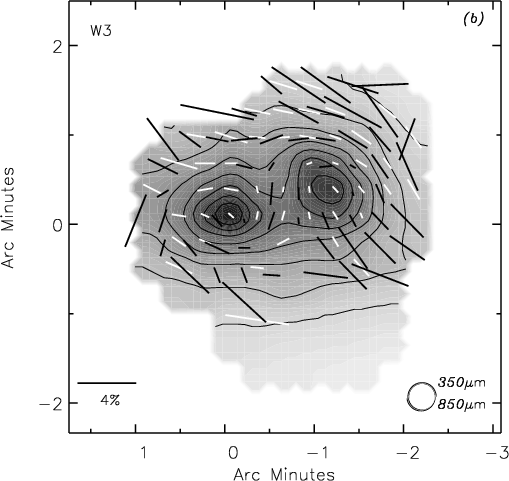}
\caption{Multiwavelength polarization vectors in the W3 molecular
  cloud overlaid on $350\,\micron$ intensity grayscale and contours.
  (\emph{a}) $60\,\micron$ (dashed-black vectors with $22\arcsec$
  resolution), $100\,\micron$ (solid-black; 35\arcsec), and
  $350\,\micron$ (solid-white; 20\arcsec). (\emph{b}) $350\,\micron$
  (solid-white vectors with $20\arcsec$ resolution) and $850\,\micron$
  (solid-black; 18\arcsec). The $350\,\micron$ polarization and
  intensity data in (\emph{a}) and (\emph{b}) are from the same
  dataset \citep{hertzarchive}.  The 60 and 100 $\micron$ data are
  from \citet{archivestokes} and the $850\,\micron$ data are from
  \citet{scubaarchive}.
}
\label{fig-w3}
\end{figure}

Observations of FIR--MM polarimetry have generated polarization maps
which span wavelengths of 60 -- 1300 $\micron$.  Figure \ref{fig-w3}
shows one such example in the W3 molecular cloud where data are
available at 60, 100, 350, and 850 $\micron$. The position angle shows
little-to-no change with wavelength in the outer portions of the
cloud.  However, in the inner regions, position angle changes are
clearly evident.  Such changes with wavelength indicate the existence
of both a changing magnetic field, as well as changes in the dust
emission properties, along the line-of-sight through the cloud
\citep{w3}.

\begin{figure}[!tb]
\centering
\includegraphics[height=1.85in]{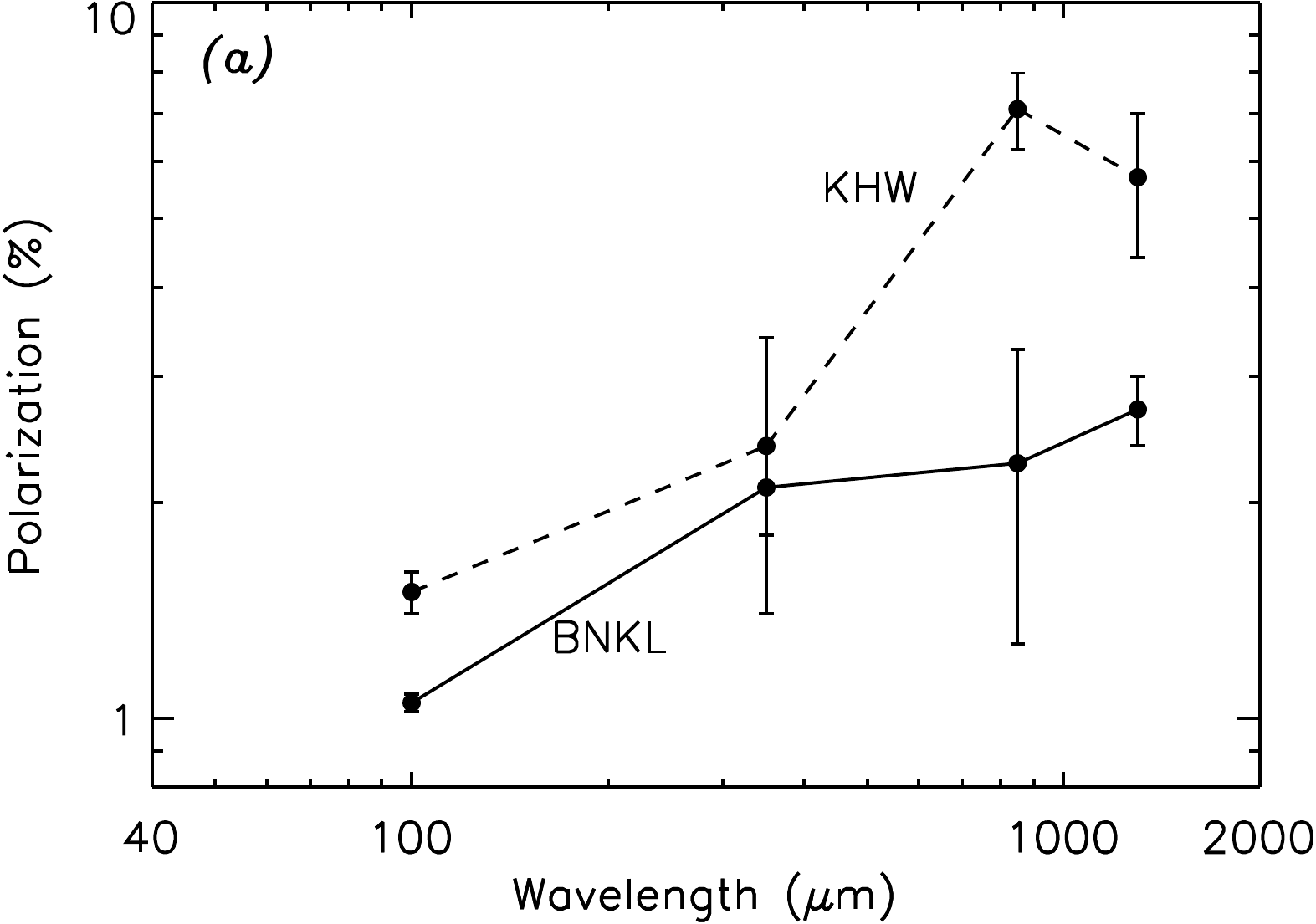}
\includegraphics[height=1.85in]{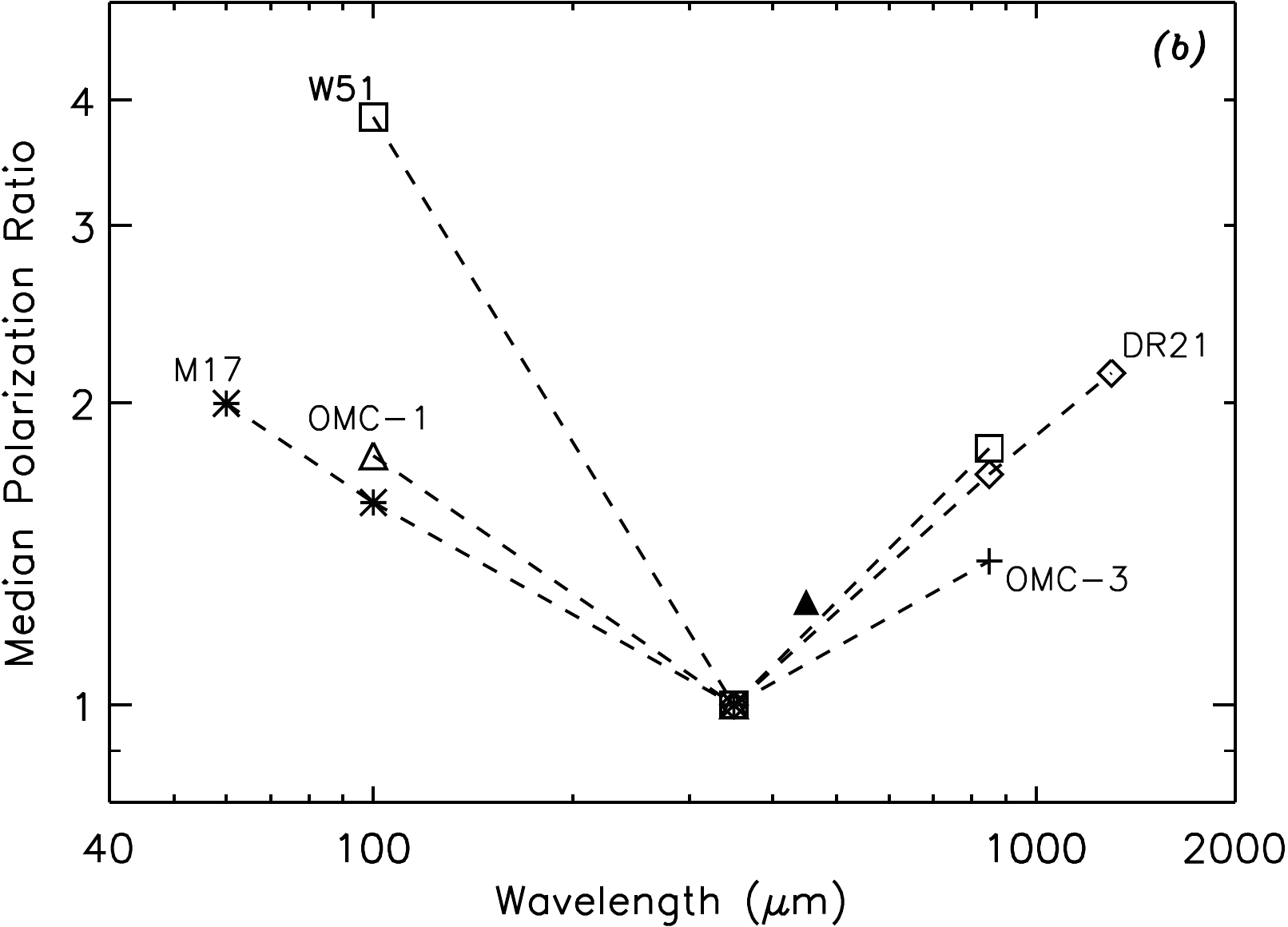}
\caption{FIR--MM polarization spectra in Galactic molecular
  clouds. (\emph{a})~Two separate cloud cores in the Orion Molecular
  Cloud (data from \citealt{dasth} and
  \citealt{coppin2000}). (\emph{b}) Cloud ``envelopes'' (away from
  dense cores) in several different clouds \citep{omc1sharp}. Envelope
  polarizations have been normalized at $350\,\micron$ within each
  cloud.  The solid triangle at $450\,\micron$ is from the data shown
  in Figure \ref{fig-omc1}\emph{b}.}
\label{fig-pspec}
\end{figure}

Changes in the polarization \emph{amplitude} with wavelength are also
observed in the FIR--MM (e.g., \citealt{pspec,mythesis,paris}).  In
the densest cores of molecular clouds the spectrum increases with
wavelength (Fig.\ \ref{fig-pspec}\emph{a}). In this case, the rise is
consistent with an opacity effect \citep{dasth}. As the opacity
increases towards shorter wavelengths the emitted polarization must
decrease, approaching zero as the emission becomes optically thick.

In cloud envelopes, where the emission is typically optically thin,
the spectrum falls with wavelength below $350\,\micron$, but rises at
longer wavelengths (Fig.\ \ref{fig-pspec}\emph{b}).  This behavior is
not consistent with a simple isothermal dust model but requires
multiple grain populations where each population's polarization
efficiency is correlated with either the dust temperature or spectral
index \citep{pspec,paris}.

\begin{figure}[!tb]
\centering
\includegraphics[height=3.17in]{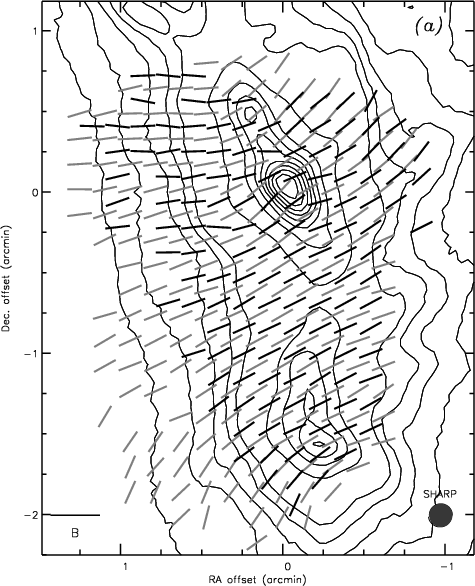}
\includegraphics[height=3.17in]{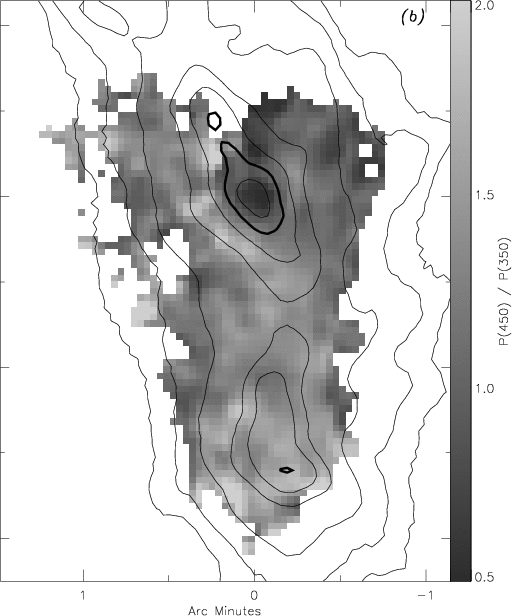}
\caption{Multi-wavelength polarimetry of the Orion Molecular Cloud
  (OMC-1) from the SHARP polarimeter \citep{omc1sharp}.  The
  $350\,\micron$ intensity peak (contours) at the coordinate origin is
  towards the Kleinmann-Low nebula. (\emph{a}) 350 (gray) and 450
  (black) $\micron$ inferred magnetic field vectors.  These vectors
  are drawn at a constant length that is not proportional to the
  polarization amplitude. (\emph{b}) The grayscale indicates the ratio
  of the 450-to-350 $\micron$ polarization amplitudes.}
\label{fig-omc1}
\end{figure}

The measured polarization spectrum is sparsely sampled in terms of the
number of wavelengths and has not been measured in very many clouds.
The goal of current observations is to better populate this
spectrum. \cite{hertzscuba08} have compared the polarization in 15
clouds at wavelengths of 350 and 850 $\micron$.  A preliminary
analysis is consistent with the results of Figure
\ref{fig-pspec}\emph{b} in which $P(850) > P(350)$ in each of these
clouds. \citet{omc1sharp} have also begun a campaign to map the
350-to-450~$\micron$ polarization ratio in the same set of Galactic
clouds.  Results in the Orion Molecular Cloud (Fig.\ \ref{fig-omc1})
show that the $P(450) / P(350)$ ratio varies across the cloud.  The
ratio is less than unity towards the SMM intensity peak of the
Kleinmann-Low nebula.  However, in the cloud envelope outside of this
core the ratio is larger, with a median of $P(450)/P(350) = 1.3$; this
is consistent with the previous measurements in other clouds shown in
Figure \ref{fig-pspec}\emph{b}.

\subsection{Cloud and Dust Models}

The rising polarization spectrum at $\lambda > 350\,\micron$ can be
reproduced with a model in which the colder grains are better aligned
than the warmer grains.  That is, for a two component model we require
temperatures $T_A > T_B$ and polarizations $p_A < p_B$\@.
\citet{bethell07} have shown that this can be achieved by applying the
radiative torque model of grain alignment
\citep[e.g.,][]{draine96,draine97} to starless clouds.  In their model
the cloud structure is ``clumpy'' (or fractal) such that external
photons can penetrate deep into the cloud.  These photons heat all
grains, but the larger grains tend to be cooler as they are more
efficient emitters.  At the same time, the alignment mechanism is more
efficient at aligning the larger grains \citep{cho05}.  Therefore,
this model predicts that the cooler grains are better aligned and that
the polarization spectrum rises with wavelength.

While the \citet{bethell07} model predicts the qualitative behavior in
part of the observed polarization spectra, it does not predict the
fall in polarization with wavelength for $\lambda < 350\,\micron$.
Additionally, their predicted spectrum rises by a factor of a few from
100 to 500 $\micron$ while the observed spectra rise within the range 350
-- 1300 $\micron$.
In real molecular clouds there exist embedded stars that provide
an additional source of photons, which will both heat and align dust
grains.  One can expect that grains closer to these stars will be
warmer and better aligned than grains that are either further from
stars or shielded from photons in optically thick clumps.  This
naturally produces grain populations in which the warmer grains are
better aligned.  The result is a polarization spectrum that falls
with wavelength.  The observed polarization spectrum with a minimum
between 100 and 850 $\micron$ can in fact be modeled by incorporating
embedded stars into the models of starless cores (A. Lazarian, private
communication).

\subsection{Observational Tests: The Spectral Energy Distribution}

\begin{figure}[!tb]
\centering
\includegraphics[width=1.0\textwidth]{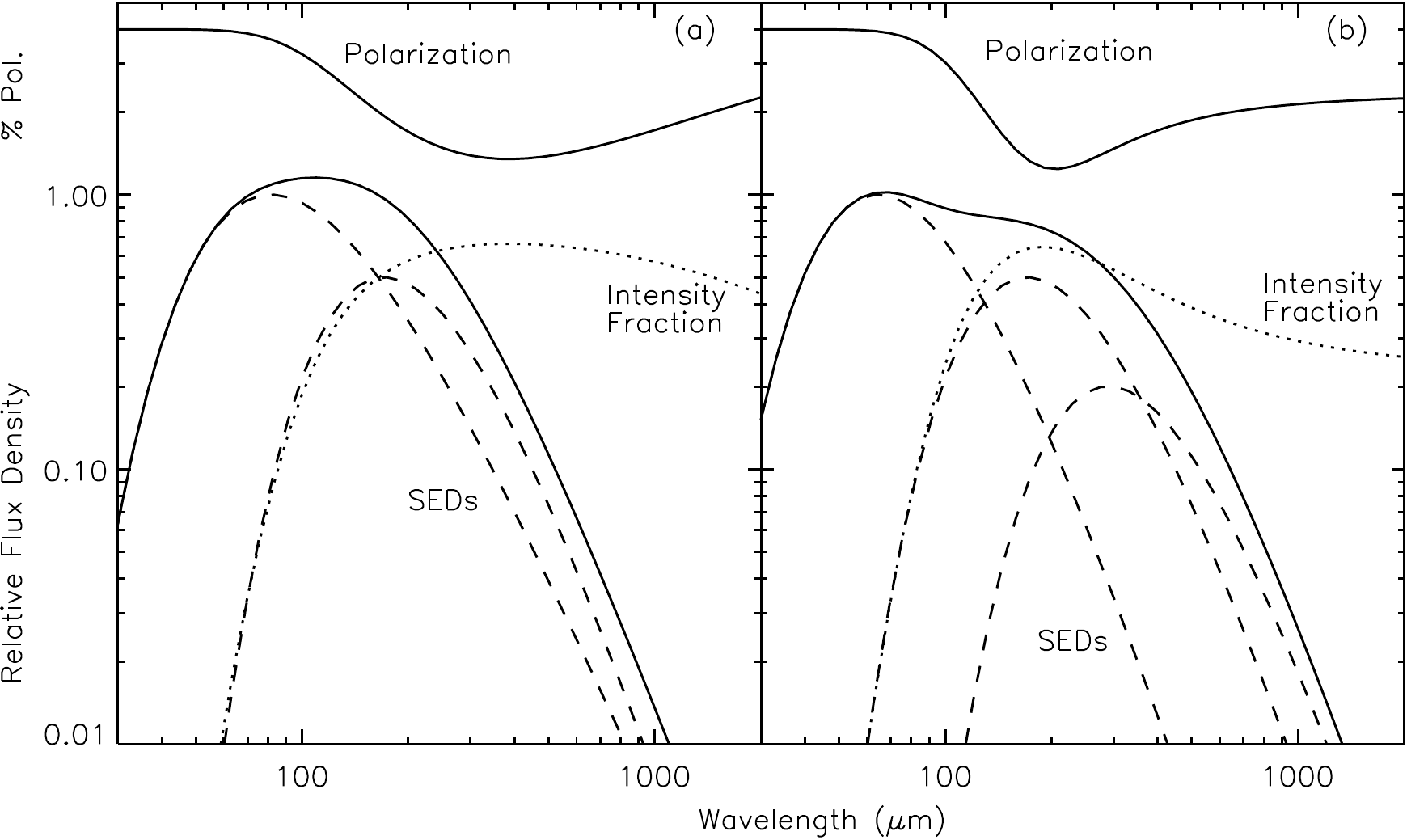}
\caption{Example SEDs and predicted polarization spectra.  Solid lines
  show the total emission or polarization from the individual
  components drawn as dashed lines. Dotted lines show the relative
  contribution to the total intensity of the unpolarized components;
  the cold component in (\emph{a}) or the intermediate temperature
  component in (\emph{b}). (\emph{a}) A two-temperature component SED
  with $T_{1,2} = 45$ and 17 K; $p_{1,2} = 4$ and 0 \%; and
  $\beta_{1,2} = 1$ and 2, respectively. (\emph{b}) A
  three-temperature component SED with $T_{1,2,3} = 45$, 17, and 10 K;
  $p_{1,2,3} = 4$, 0, and 3 \%, respectively; and $\beta_{1,2,3} = 2$
  for all components.}
\label{fig-seds}
\end{figure}

If multiple temperatures or spectral indices exist along the line of
sight in molecular clouds, as predicted by the model of the
polarization spectrum, then one expects this to have an observable
effect on the total intensity spectral energy distribution (SED\@).
Unfortunately, little data exist in the clouds and wavelength range
($\lambda \sim 50 - 1000$ $\micron$) of interest to perform definitive
tests using measurements of both the polarization spectrum and SEDs.
\citet{mythesis} found that existing SEDs in the OMC-1 molecular cloud
were consistent with the multiple temperature hypothesis.  However,
the SEDs were sparsely sampled (maps of the region were available at
only 6 wavelengths from 60 to 1100 $\micron$) and the flux
uncertainties were large (calibration uncertainties as large as 30\%).
As a result it was difficult to distinguish between fits to
one-temperature component models vs.\ two-temperature component
models.

Consider the multi-component SEDs in Figure \ref{fig-seds}.  The
intensity of each component $i$, at frequency $\nu$, is given by
$I_\nu(T_i) = N_i \; \nu^{\beta_i} \; B_\nu(T_i)$, where $N_i$ is the
column density of each component. The polarization spectrum is high at
wavelengths where the total intensity is dominated by a warm polarized
component and low at wavelengths dominated by an unpolarized
component.  In the two-component model with different spectral indices
(Fig.\ \ref{fig-seds}\emph{a}) a polarization minimum occurs because
the polarized component dominates at both short and long wavelengths,
but not intermediate wavelengths.  In the three-component model with
equal spectral indices (Fig.\ \ref{fig-seds}\emph{b}) the polarization
minimum occurs because the intermediate temperature is unpolarized
while the colder and warmer components are both polarized.  The latter
model is similar to the \citet{bethell07} model but with the addition
of embedded stars.

From these models we see that the wavelength of the polarization
minimum is dependent only on the spectral shape of the components and
independent of both the relative column densities $N_i$ and
polarization efficiencies $p_i$. For example, as any temperature $T_i$
is changed the corresponding normalized SED will simply shift along
the wavelength axis, changing the location at which it contributes its
maximum to the total intensity.  One could also shift the curves in
intensity by changing the relative amounts of cold and warm dust.
However, such intensity shifts will only change the absolute and
relative values of the polarization, not its spectral shape.  As a
result, one could use the polarization spectrum to further constrain
models of the SED \citep{aod}.

\subsection{Observational Tests: Embedded Stars}

\begin{figure}[!tb]
\centering
\includegraphics[height=0.6\textwidth]{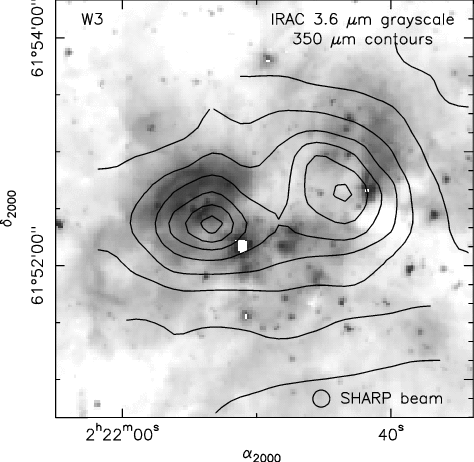}
\caption{The W3 molecular cloud at $3.6\,\micron$ from
  \emph{Spitzer/IRAC} (grayscale; courtesy C. D. Dowell) and
  $350\,\micron$ from Hertz (contours; \citealt{w3}).  The
  $350\,\micron$ intensity and polarization measurements in Figure
  \ref{fig-w3}, with a $20\arcsec$ spatial resolution, are unable to
  resolve individual stars; future polarimeters with better spatial
  resolution (10\arcsec; \citealt{sharpao,spie2007}) are expected to
  do so.}
\label{fig-w3stars}
\end{figure}

Another test of the polarization spectrum model is to directly compare
the locations of embedded stars in molecular clouds with the measured
polarization amplitude.  If photons are required for efficient grain
alignment \cite[e.g.,][]{draine97} then one should find increased
alignment efficiency (and therefore increased polarization) near
embedded stars when compared to nearby locations without stars.  While
embedded stars are readily found using near-infrared imaging, such
comparisons are difficult. For example, Figure \ref{fig-w3stars} compares
$3.6\,\micron$ emission (mostly stellar) and $350\,\micron$ emission
(cold dust) in the W3 molecular cloud.  The $20\arcsec$ resolution
(typical in the SMM) of the $350\,\micron$ observations is
insufficient to resolve individual stars due to the relatively high
stellar density.  Also shown is the $10\arcsec$ beam of the SHARP
polarimeter, which is capable of resolving individual stars in the
outer regions of the map.

\newcounter{number} \setcounter{number}{2}
\newcommand\hii{H$\;${\small\Roman{number}}\ } FIR
instruments with higher spatial resolution ($\sim 5\arcsec$;
\citealt{spie2007,hale}) will likely be needed for a more unambiguous
comparison in the densest regions.  FIR observations will also
be more sensitive to the warm dust near these embedded stars, perhaps
providing a larger contrast between the cold and warm dust in these
regions.  For example, \citet{w3} have shown that $60\,\micron$
polarization observations towards a radio \hii region in W3 show a
clear increase as opposed to much lower polarization away from the
\hii region.  This same trend is not obvious at $350\,\micron$ towards
the same feature.

\section{Summary}

The increasingly large database of far-infrared and submillimeter
polarimetry is just beginning to allow studies of interstellar
magnetic fields across a wide range of size scales.  Future
polarization measurements with interferometers, such as the
Submillimeter Array \citep{sma2008} and
ALMA\footnote{{http://www.alma.nrao.edu/}} \citep{alma2008}, will
continue to sample star-forming regions on arc-second scales.
Single-dish observations (e.g., \citealt{sharpao,bastien08}) will be
required to detect the extended emission from nascent molecular clouds
and connect the smallest-scale fields to intermediate scales.  Mapping
magnetic fields on the larger scales (several arc-minutes -- several
degrees) requires large beams which must necessarily sacrifice spatial
resolution (e.g., \citealt{novak03,li06}).  Current large-scale
(sub-)millimeter surveys designed to measure the polarization of the
cosmic microwave background (e.g., \citealt{page2007,bicep2008}) on
degree-scale sizes are too large to perform direct comparisons with
the existing arc-minute scale observations.  However, future survey
experiments such as the \emph{Planck} satellite
\citep{tauber04,aumont08} are expected to reach resolutions of a few
arcminutes.

The existing data also allow for further studies of the dependence of
the polarization amplitude on wavelength.  The observed spectrum with
a polarization minimum near $\lambda \sim 350\,\micron$ can be
understood if there exist correlations between the polarization and
either the dust temperature or emissivity.  Such correlations are
expected to occur naturally in molecular clouds given realistic
simulations \citep[][]{bethell07} which incorporate modern grain
alignment theory \citep[e.g.,][]{alexreview2}.

However, the data in both polarization and total intensity are too
sparse to provide quantitative tests of the polarization spectrum
models.  Therefore, our immediate goal is to collect more data in
terms of increased wavelength coverage, different types of cloud
environments (temperatures and densities), and increased spatial
resolution.  Future work in all these areas will continue to come from
a number of submillimeter instruments on both single-dish telescopes
and interferometric arrays.  Data in the far-infrared will also
provide both increased spatial resolution and access to shorter
wavelengths.  These wavelengths are crucial to further test and
characterize the polarization spectrum and will require far-infrared
instruments on airborne or space-based observatories.

\acknowledgements I would like to thank the conference organizers for
inviting me to present this work.  Many collaborators have contributed
to this work over the years, including Roger Hildebrand, Darren
Dowell, Larry Kirby, Giles Novak, Martin Houde, Michael Attard,
Hua-bai Li, Megan Krejny, Jessie Dotson, Jackie Davidson, Brenda
Matthews, and Hiroko Shinnaga.  The author has received support from
the U.S. National Science Foundation award AST 05-40882 to the
California Institute of Technology.

\bibliography{hildebrand,me,polar,microwave,dust,theory,instrument,other,tmp}
\bibliographystyle{apj}

\end{document}